\documentclass[twocolumn,aps,pre,showpacs,amsmath,amssymb]{revtex4-1}
\usepackage{epsfig}
\usepackage{graphicx}
\usepackage{dcolumn}
\usepackage{bm}
\usepackage[colorlinks=true,dvipdfm]{hyperref}

\begin{document}

\title{Scaling theory of entanglement entropy in confinements near quantum critical points}
\author{Xuanmin Cao, Qijun Hu}
\author{Fan Zhong}
\thanks{Corresponding author. E-mail: stszf@mail.sysu.edu.cn}
\affiliation{State Key Laboratory of Optoelectronic Materials and Technologies, School of
Physics, Sun Yat-sen University, Guangzhou 510275, People's
Republic of China}
\date{\today}

\begin{abstract}
We propose a unified scaling theory of entanglement entropy in the confinements of finite bond dimensions, dynamics and system sizes. Within the theory, the finite-entanglement scaling introduced recently is generalized to the dynamics subjected to a linear driving along with a finite system size. Competition among the three scales as well as the correlation length of the system is analysed in details. Interesting regimes and their complicated crossovers together with their characteristics follow naturally. The theory is verified with the one-dimensional transverse-field Ising model under a linear driving.
\end{abstract}

\maketitle
\section{\label{intro}Introduction}
As the weirdest aspect of quantum mechanics, entanglement entropy has attracted great concern and has served as an important tool to characterize quantum phase transitions of many-body systems~\cite{Osterloh,Osborne,GVidal,JVidal,Calabrese04,Wu,Refael,Orus,Laflorencie,Brandao,Li,Chiara,Zhao,Chandran,Lundgren,Braganco,Amico,Eisert} including in particular topological order~\cite{Preskill,Wen,Pollmann1,Jiang}.
For a system described by a pure state $|\psi\rangle$, one can divide the system into two parts with their respective corresponding orthonormal bases $\{|\psi^{\rm L}_k\rangle\}$ and $\{|\psi^{\rm R}_k\rangle\}$ such that
\begin{equation}
|\psi\rangle=\sum_k \lambda_k |\psi^{\rm L}_k\rangle\otimes |\psi^{\rm R}_k\rangle
\label{schdeco}
\end{equation}
with the finite coefficients $\lambda_k\geq 0$. This is the Schmidt decomposition.
$|\psi^{\rm L}_k\rangle$ and $|\psi^{\rm R}_k\rangle$ are just the eigenvectors of the reduced density operators $\rho_{L/R}={\rm tr}_{R/L}|\psi\rangle\langle\psi|= \sum_k \lambda_k^2 |\psi_{L/R,k}\rangle\langle\psi_{L/R,k}|$.
$\rho_{L}$ and $\rho_{R}$ thus share a mutual spectrum of $\lambda_k^2$ satisfying $\sum_k \lambda_k^2=1$ and thus are equally mixed.
As a result, by measuring the mixedness of the reduced density operators, one can obtain the entanglement~\cite{Amico}.

Among the various measures of the mixedness of the pure state, bipartite entanglement entropy is a proper measure. It is defined as the von Neumann entropy of either part of the reduced density matrices,
\begin{equation}
S= -\sum_k \lambda_k^2\ln\lambda_k^2.
\label{Sdefine}
\end{equation}
Accordingly, if and only if a pure state is the direct product of the pure states of the two parts, viz., only one $\lambda_k$ is finite, it is not entangled and $S=0$.

The entanglement entropy possesses a universal specific form for a quantum phase transition in a one-dimensional system of length $L$. Near a quantum critical point with a correlation length $\xi\ll L$, the bipartite entanglement entropy is given asymptotically by~\cite{Calabrese}
\begin{equation}
S\sim\frac{c}{6}\ln \xi,
\label{etoxi}
\end{equation}
where $c$ is the central charge. This divergence of the entanglement entropy with $\xi$ implies that an unbounded number of states are needed to characterize the entanglement of a state near the critical point.

However, there exist situations of confinement in which the divergence of the entanglement entropy is limited. One case is when $\xi$ gets longer than $L$ so that the latter ought to impose limit on the entanglement. Indeed, the entanglement entropy of one half of the system is now restricted by $L$ and behaves as~\cite{Holzhey,GVidal,Calabrese}
\begin{equation}
S\sim\frac{c}{6}\ln L
\label{etol}
\end{equation}
to the leading order in $L$. This can be understood reasonably as a finite size effect in which the length scale is now dominated by $L$ instead of $\xi$. Confined as it is, $S$ as given in Eq.~(\ref{etol}) can be employed in a finite-size scaling (FSS) analysis~\cite{Cardy} to access critical properties~\cite{Berkovits,Zhao,Chen}.

There exists another confinement to the entanglement entropy. This is the confinement in which only a limited number of states are available to support the entangled state. Such a confinement leads to the so called finite-entanglement scaling (FES)~\cite{Tagliacozzo}. As exact solutions of quantum systems are often difficult or even impossible to acquire, appropriately approximated approaches on the basis of matrix product states (MPS)~\cite{Klumper,Verstraete1,Schollwock} are often introduced. One of the ramifications of the MPS form, the so-called infinite time-evolving block decimation (iTEBD) algorithm\cite{Vidalitebd}, can even well simulate an infinite chain. However, it has been found that~\cite{Tagliacozzo} only a certain amount of entanglement is captured at criticality. This is caused by the finite bond dimension $D$, also known as a truncation, which is the dimension of the matrices and thus the number of states kept. The finite bond dimension also limits the accuracy of the density matrix renormalization group algorithm near critical points~\cite{Andersson}. Upon assuming an effective correlation length $\xi_D$ given asymptotically by
\begin{equation}
\xi_D\sim D^{\kappa}\label{xid}
\end{equation}
with a universal exponent $\kappa$, the entanglement entropy now exhibits
\begin{equation}
S\sim\frac{c}{6}\ln D^{\kappa}.
\label{etod}
\end{equation}
An FES was then proposed in analogy to FSS and applied to estimate the critical point and the critical exponents~\cite{Tagliacozzo}. The exponent $\kappa$ was later found analytically to be related to the central charge through
\begin{equation}
\kappa=\frac{6}{c(\sqrt{12/c}+1)}
\label{kappa}
\end{equation}
rather than to the scaling dimension of an operator~\cite{Pollmann0}. It is suggested that both the behaviors of FES and FSS are a direct consequence of conformal invariance~\cite{Pollmann0,Blote}. The relation between FES and FSS was studied in Ref.~\cite{Pirvu}. It was also concluded that the former has less computational cost than the latter~\cite{Nagy}.

Recent experimental advances in manipulating the real-time evolution of ultracold atoms~\cite{Greiner,Kinoshita,Hofferberth,Zhang} have stimulated a resurgence in studying the dynamics of continuous phase transitions~\cite{Dziarmaga,polrmp,inexper4,Zhong1,Zhong2,Yin,Yin2,Liu,Huang,Huang2,Pelissetto}. On the one hand, when a one-dimensional chain is quenched abruptly to a new Hamiltonian and then relaxes, the entanglement entropy is found to increase linearly with time up to a saturation value~\cite{Cala,Scha,Moore}; while in the imaginary-time relaxation, in which the real time of the Schr\"{o}dinger equation for the state evolution is replaced by an imaginary time so that the evolution is nonunitary, the entanglement entropy increases as the logarithm of time at the critical point~\cite{Yin3}. On the other hand, when a chain is driven with a finite temporal rate through the critical point, the entanglement entropy is proportional to the logarithm of the driving rate~\cite{Cincio,Moore,Canovi}. Besides the entanglement entropy, the entanglement spectrum~\cite{Li,Chiara,Chandran,Lundgren,Braganco}, a form of the Schmidt eigenvalues $\lambda_k$, has also been studied in both driving and relaxational dynamics~\cite{Torlai,Canovi,Hu}. A finite-time scaling (FTS)~\cite{Zhong1,Zhong2} theory for the Schmidt gap, the difference between the two largest Schmidt eigenvalues, has been verified and employed to estimate critical properties and its relation with FSS has also been studied~\cite{Hu}.

FTS describes the universal dynamics of a system that is driven with a finite rate $R$ through its critical point~\cite{Zhong1,Zhong2}. The basic idea is that the driving introduces a finite time scale~\cite{Zhong1,Zhong2,Yin,Huang}
\begin{equation}
\zeta_R\sim R^{-z/r},\label{zd}
\end{equation}
where $z$ is the dynamic critical exponent and $r$ is the renormalization-group eigenvalue of $R$. It depends on which parameter of the system is being varied with time and is related to $z$ and the scaling dimension of that parameter~\cite{Zhong1,Zhong2}. When the correlation time of the system, $\zeta$, is smaller than $\zeta_R$, the system can adiabatically follow the driving and the driving itself is only a small perturbation. In the extremely reverse case of $\zeta_R\ll\zeta$, however, the system falls out of equilibrium and its behavior is determined by $R$. This is the FTS regime, which is similar to the static FSS regime in the case of $L\ll\xi$. FTS describes the universal scaling behaviors of both the adiabatic and nonequilibrium regimes and their crossover well and can also be applied to determine the critical properties similar to FSS~\cite{Zhong1,Zhong2,Yin,Yin2,Liu,Huang,Hu,Pelissetto}. It can even be applicable to the case in which the system is driven from near to its critical point with a nonequilibrium initial state, the state which gives rise to the so-called initial slip~\cite{Janssen}, resulting in the competition of FTS and the initial slip~\cite{Huang2}.

The finite time scale, Eq.~(\ref{zd}), corresponds to a driven-imposed finite length scale~\cite{Zhong2}
\begin{equation}
\xi_R\sim\zeta_R^{1/z}\sim R^{-1/r},\label{xir}
\end{equation}
similar to the relation between the correlation time and the correlation length $\zeta\sim\xi^z$~\cite{Cardy}. Competition between FTS and FSS can then be envisioned by comparing $\xi_R$ and $L$~\cite{Zhong2,Liu,Huang,Hu}. The logarithmic dependence of the entanglement entropy on $R$ can then be understood as a dynamic confinement of $S$ and can be obtained by replacing $\xi$ with $\xi_R$ in Eq.~(\ref{etoxi}), leading to~\cite{Cincio,Moore,Canovi}
\begin{equation}
S\sim\frac{c}{6}\ln R^{-1/r}.
\label{etor}
\end{equation}
However, as pointed out above, this is true only within the FTS regime, where the system is dominated by $R$.

Here we generalize the FES to dynamics and study the scaling of the entanglement entropy in confinements of the bond dimension, the system size, and the driving dynamics together near a quantum critical point. A unified scaling theory that takes into account all the four length scales $\xi$, $\xi_D$, $L$, and $\xi_R$ or the corresponding time scales is proposed. Competition among all the length scales can then be studied in details. From the theory, the shortest among the four long scales controls the behavior of the system, which then falls in a regime dominated by the scale. All the remaining scales only contribute as perturbations. If the relative lengths of the scales change, occurs a crossover to a new regime governed by the new shortest scale. Accordingly, the relations of the entanglement entropy in Eqs.~(\ref{etoxi}), (\ref{etol}), (\ref{etod}), and (\ref{etor}) are only valid in their respective regimes, while outside the regimes they are saturated to values that depend on those of the other parameters. In particular, FES works only when $R$ is small enough and conversely FTS emerges when $R$ is sufficiently large, both on condition that $L$ and $\xi$ are large. Moreover, owing to the competition of several scales, complicated crossovers emerge. All the results are born out by numerical results from the one-dimensional transverse field Ising model. Our theory can be extended straightforwardly to take into account other length or time scales.

In the following, we first present the scaling theory of the entanglement entropy for the competition of different length or time scales in Sec.~\ref{theory}. Various limited forms in which some scales are long enough to serve as perturbations are recovered. Crossovers between different scales are studied in details. Some special loci are also presented for numerical test of the theory. Then, after the introduction of the model and numerical algorithm in Sec.~\ref{model}, we present the numerical results to confirm our theory in Sec.~\ref{numr}. Finally, a summary is given in Sec.~\ref{sum}.

\section{\label{theory}Scaling theory of the entanglement entropy in confinements}
In this section, we will develop a unified theory for the scaling of the entanglement entropy in the various confinements discussed in Sec.~\ref{intro}. The theory generalizes those of FTS, FES, and FSS by combining all the relevant length scales or time scales together consistently and can thus account for their competition.

As the universal form of the entanglement entropy is determined by the correlation length, we first study the scaling of the correlation length in Sec.~\ref{correlation} and then apply it to the entanglement entropy in Sec.~\ref{entropy}. Finally, some special loci of the scaling theory are presented in Sec.~\ref{specialloci}.

\subsection{\label{correlation}Scaling of the correlation length}
Let $g$ be the distance to a quantum critical point and change linearly with time as
\begin{equation}
g=Rt\label{grt}
\end{equation}
for a constant rate $R$. Our scaling theory is based on the following ansatz for the correlation length
\begin{equation}
\xi(g,D^{-1},L^{-1},R)=b\xi(gb^{1/\nu},D^{-1}b^{1/\kappa},L^{-1}b,Rb^r),
\label{xitob}
\end{equation}
under a length rescaling of factor $b$, where $\nu$ is the correlation length critical exponent. In Eq.~(\ref{xitob}), we have neglected dimensional factors for simplicity. Because time rescales as $tb^{-z}$, Eq.~(\ref{grt}) then results in~\cite{Zhong1,Zhong2}
\begin{equation}
r=z+1/\nu.\label{rznu}
\end{equation}

One can reckon from the ansatz, Eq.~(\ref{xitob}), the various length scales mentioned above. For example, choosing a scale such that $gb^{1/\nu}$ is a constant, one finds the usual equilibrium correlation length near the critical point
\begin{subequations}
\label{xiscaling}
\begin{equation}
\xi=|g|^{-\nu}f_{\xi g}(D^{-1}|g|^{-\nu/\kappa},L^{-1}|g|^{-\nu},R|g|^{-r\nu}),\label{xitog}
\end{equation}
where $f_{\xi g}$ is a universal scaling function. To be consistent, the arguments of the scaling function must be vanishingly small in order to keep it analytic. This means that the equilibrium correlation length $\xi\sim|g|^{-\nu}$ is recovered for a system of sufficiently large size $L\gg\xi$, sufficiently large bond dimension $D\gg\xi^{1/\kappa}$, and sufficiently small rate $R\ll\xi^{-r}$, not necessarily for $L=\infty$, $D=\infty$, and $R=0$. These conditions for $D$ and $R$ are simply $\xi_D\gg\xi$ and $\xi_R\gg\xi$ using $\xi_D$ and $\xi_R$ from Eqs.~(\ref{xid}) and (\ref{xir}), respectively. Using Eq.~(\ref{xitob}), one obtains $\xi\sim\xi_D$, $\xi\sim \xi_R$, and $\xi\sim L$ from
\begin{eqnarray}
\xi&=&D^{\kappa}f_{\xi D}(gD^{\kappa/\nu},L^{-1}D^{\kappa},RD^{r\kappa}),\label{xitod}\\
\xi&=&R^{-1/r}f_{\xi R}(gR^{-1/r\nu},D^{-1}R^{-1/r\kappa},L^{-1}R^{-1/r}),\qquad\label{xitor}\\
\xi&=&Lf_{\xi L}(gL^{1/\nu},D^{-1}L^{1/\kappa},RL^{r}),\label{xitol}
\end{eqnarray}
\end{subequations}
respectively, similar to Eq.~(\ref{xitog}), when the other length scales are relatively large enough than the one specified, where all $f$ are universal scaling functions.
In other words, the effective correlation length is governed by the shortest length scales among those length scales that are sufficiently longer than the microscopic ones and thus exhibit universal scaling.

The four length scales correspond to four time scales in dynamics. Because $g$, $R$, and $t$ are related by Eq.~(\ref{grt}), they are not independent. As a result, one can choose arbitrarily any pair out of the trio as independent variables other than the pair $g$ and $R$ as was done in Eq.~(\ref{xitob}). Similar manipulations then lead to various time scales corresponding to their respective length scales. In particular, the finite bond dimension $D$ defines an associated finite time scale $\zeta_D\sim D^{z\kappa}$. Moreover, the shortest time scale is the dominant time scale that controls the evolution of the system similar to the spatial case.

The four equations, (\ref{xitog})--(\ref{xitor}), are the (quasi-)equilibrium scaling, the FES, the FTS, and the FSS of the correlation length, respectively, as the subscripts of the scaling functions already indicate. Each scaling shows when its corresponding length scale or time scale is shortest and dominates as pointed out above.

Nevertheless, every single equation in Eq.~(\ref{xiscaling}) can also describe other scalings besides its dominant one. As such, however, different scalings emerge as different regimes that are controlled by their leading scalings. There exists a crossover between each pair of two different regimes.
This crossover from one regime dominated by one scale to another one dominated by another scale occurs when the relative magnitude of the scales changes. For example, a crossover from the FES regime to the FTS regime occurs when $\xi_R\ll\xi_D$ and vice versus. As a consequence, the scaling functions are related. Their relations can be found by ensuring the correct form of the scaling to be crossed over to. Thus,
\begin{subequations}
\label{xicross}
\begin{eqnarray}
f_{\xi R}(X,Y,Z)&=&X^{-\nu}f_{\xi g}(YX^{-\frac{\nu}{\kappa}},ZX^{-\nu},X^{-r\nu}),\label{xirtog}\\
f_{\xi D}(X,Y,Z)&=&X^{-\nu}f_{\xi g}(X^{-\frac{\nu}{\kappa}},YX^{-\nu},ZX^{-r\nu}),\label{xidtog}\\
f_{\xi L}(X,Y,Z)&=&X^{-\nu}f_{\xi g}(YX^{-\frac{\nu}{\kappa}},X^{-\nu},ZX^{-r\nu}),\qquad\label{xiltog}\\
f_{\xi R}(X,Y,Z)&=&Y^{-\kappa}f_{\xi D}(XY^{-\frac{\kappa}{\nu}},ZY^{-\kappa},Y^{-r\kappa}),\label{xirtod}\\
f_{\xi R}(X,Y,Z)&=&Z^{-1}f_{\xi L}(XZ^{-\frac{1}{\nu}},YZ^{-\frac{1}{\kappa}},Z^{-r}),\label{xirtol}\\
f_{\xi D}(X,Y,Z)&=&Y^{-1}f_{\xi L}(XY^{-\frac{1}{\nu}},Y^{-\frac{1}{\kappa}},ZY^{-r}),\label{xidtol}
\end{eqnarray}
\end{subequations}
where $X$, $Y$, and $Z$ denote the corresponding scaled variables of the scaling functions involved. For example, $X$, $Y$, and $Z$ of $f_{\xi R}$ in Eq.~(\ref{xirtog}) are just the three scaled variables of the same function in Eq.~(\ref{xitor}). Equation~(\ref{xicross}) can also be obtained by equating the corresponding pairs of $\xi$ in Eq.~(\ref{xiscaling}). Indeed, equating Eqs.~(\ref{xitor}) and (\ref{xitog}) results in Eq.~(\ref{xirtog}). Equations~(\ref{xirtog}), (\ref{xirtol}), (\ref{xidtol}), and (\ref{xiltog}) describe the crossovers of FTS to equilibrium scaling~\cite{Zhong1,Zhong2}, FTS to FSS~\cite{Zhong2,Liu,Huang,Hu}, FES to FSS~\cite{Pirvu}, and FSS to equilibrium~\cite{Cardy}, respectively, while Eq.~(\ref{xidtog}) is the crossover of FES to equilibrium scaling. The competition between FTS and FES, Eq.~(\ref{xirtod}), is the generalization here. It implies that when $Y=D^{-1}R^{-1/r\kappa}$ gets large while the other two scaled variables remain small, the FTS scaling function $f_{\xi R}$ behaves singularly as $Y^{-\kappa}$ and thus crosses over to the FES regime, within which $\xi_D$ dominates and all scaled variables become small. The crossover from FES to FTS can be inferred by inverting Eq.~(\ref{xirtod}) or directly from Eq.~(\ref{xiscaling}) again. So can the others in Eq.~(\ref{xicross}). The results for the other relations among $R$, $D$, and $L$ are
\begin{subequations}
\label{xicrossi}
\begin{eqnarray}
f_{\xi D}(X,Y,Z)&=&Z^{-\frac{1}{r}}f_{\xi R}(XZ^{-\frac{1}{r\nu}},Z^{-\frac{1}{r\kappa}},YZ^{-\frac{1}{r}}),\label{xidtor}\\
f_{\xi L}(X,Y,Z)&=&Z^{-\frac{1}{r}}f_{\xi R}(XZ^{-\frac{1}{r\nu}},YZ^{-\frac{1}{r\kappa}},Z^{-\frac{1}{r}}),\qquad\label{xiltor}\\
f_{\xi L}(X,Y,Z)&=&Y^{-\kappa}f_{\xi D}(XY^{-\frac{\kappa}{\nu}},Y^{-\kappa},ZY^{-r\kappa}).\label{xiltod}
\end{eqnarray}
\end{subequations}

Equations~(\ref{xicross}) and (\ref{xicrossi}) are different from the previous studies of similar crossovers because there are now more variables. This results in complicated behaviors in the crossovers.
For example, in the absence of $D$, e.g., when $D\to\infty$, the crossover from the FTS to the FSS, Eq.~(\ref{xirtol}), is quite simple. As $Y=0$, at the critical point at which $g=0$, the scaling function $f_{\xi R}$ becomes a function of a single variable. In a double logarithmic plot, $f_{\xi R}$ versus $Z$ changes from a horizontal line in the FTS regime to an inclined one with a slope of $-1$ in the FSS regime. This is in fact what has been demonstrated in Refs.~\cite{Liu,Huang,Hu}, in which the only difference is that observables other than $\xi$ are studied and thus their slopes in the FSS regimes are different. However, in the presence of $D$, even at the critical point at which $g=0$, the scaling functions are surfaces in three-dimensional spaces. Their behavior will be analyzed in the following in association with the entanglement entropy that we numerically study in this paper.

\subsection{\label{entropy}Scaling of the entanglement entropy}
From the ansatz~(\ref{xitob}), the universal form of the entanglement entropy near the quantum critical point, Eq.~(\ref{etoxi}), now assumes
\begin{equation}
S(g,D^{-1},L^{-1},R)=\frac{c}{6}\ln b+\frac{c}{6}S(gb^{1\over\nu},D^{-1}b^{1\over\kappa},L^{-1}b,Rb^r)
\label{ee}
\end{equation}
for a state on a finite support in a finite system under driving. We can then choose various scales to arrive at the various scaling forms of $S$ similar to what has been done for $\xi$. Of course, we can directly insert Eq.~(\ref{xiscaling}) into Eq.~(\ref{etoxi}) to reach identical results. In the following, we just list them.

In correspondence to Eq.~(\ref{xiscaling}), the equilibrium scaling, the FES, the FTS, and the FSS of the entanglement entropy are
\begin{subequations}
\label{sscaling}
\begin{eqnarray}
S&=&\frac{c}{6}\ln|g|^{-\nu}+f_{Sg}(D^{-1}|g|^{-\frac{\nu}{\kappa}},L^{-1}|g|^{-\nu},R|g|^{-r\nu}),\nonumber\\\label{stog}\\
S&=&\frac{c}{6}\ln D^{\kappa}+f_{SD}(gD^{\frac{\kappa}{\nu}},L^{-1}D^{\kappa},RD^{r\kappa}),\label{stod}\\
S&=&\frac{c}{6}\ln R^{-\frac{1}{r}}+f_{SR}(gR^{-\frac{1}{r\nu}},D^{-1}R^{-\frac{1}{r\kappa}},L^{-1}R^{-\frac{1}{r}}),\nonumber\\\label{stor}\\
S&=&\frac{c}{6}\ln L+f_{SL}(gL^{\frac{1}{\nu}},D^{-1}L^{\frac{1}{\kappa}},RL^{r}),\label{stol}
\end{eqnarray}
\end{subequations}
respectively, where the scaling functions for $S$ are related to those for $\xi$ by
\begin{equation}
f_{Si}=\frac{c}{6}\ln f_{\xi i}\label{fsi}
\end{equation}
for $i=g$, $D$, $R$ and $L$. Equation~(\ref{sscaling}) recovers as expected Eqs.~(\ref{etoxi}), (\ref{etod}), (\ref{etor}), and (\ref{etol}), respectively, when the corresponding scaled variables in their scaling functions vanish. When the scaled variables take on small finite values, the scaling functions then constitute subleading contributions to the leading behaviors.

When one of the scaled variable gets large, the corresponding scale takes over and becomes the dominant scale and thus a crossover to its dominating regime occurs. The crossovers for the three scales relating to $R$, $L$, and $D$ are now
\begin{subequations}
\label{scross}
\begin{eqnarray}
f_{SD}&=&\frac{c}{6}\ln Z^{-\frac{1}{r}}+ f_{SR}(XZ^{-\frac{1}{r\nu}},Z^{-\frac{1}{r\kappa}},YZ^{-\frac{1}{r}}),\label{sdtor}\\
f_{SL}&=&\frac{c}{6}\ln Z^{-\frac{1}{r}}+ f_{SR}(XZ^{-\frac{1}{r\nu}},YZ^{-\frac{1}{r\kappa}},Z^{-\frac{1}{r}}),\qquad\label{sltor}\\
f_{SD}&=&\frac{c}{6}\ln Y^{-1}+ f_{SL}(XY^{-\frac{1}{\nu}},Y^{-\frac{1}{\kappa}},ZY^{-r}),\label{sdtol}\\
f_{SR}&=&\frac{c}{6}\ln Z^{-1}+ f_{SL}(XZ^{-\frac{1}{\nu}},YZ^{-\frac{1}{\kappa}},Z^{-r}),\label{srtol}\\
f_{SR}&=&\frac{c}{6}\ln Y^{-\kappa}+ f_{SD}(XY^{-\frac{\kappa}{\nu}},ZY^{-\kappa},Y^{-r\kappa}),\label{srtod}\\
f_{SL}&=&\frac{c}{6}\ln Y^{-\kappa}+ f_{SD}(XY^{-\frac{\kappa}{\nu}},Y^{-\kappa},ZY^{-r\kappa}),\label{sltod}
\end{eqnarray}
\end{subequations}
as can be found from the methods for $\xi$ or from Eqs.~(\ref{xicross}), (\ref{xicrossi}), and (\ref{fsi}), where we have dropped the symbolically-identical arguments $(X, Y, Z)$ for all the scaling functions on the left-hand sides for simplicity. One can convince oneself that the logarithmic terms in Eq.~(\ref{scross}) just cancel the originals and produce the new leading singularities in Eq.~(\ref{sscaling}) correctly. Therefore, $f_{SD}$ versus $\ln Z=\ln(RD^{r\kappa})$ is a horizontal line in the FES regime, in which $X=gD^{\kappa/\nu}$ and $Y=L^{-1}D^{\kappa}$ as well as $Z=RD^{r\kappa}$ are all vanishingly small. From Eq.~(\ref{sdtor}), it then changes to an inclined line of a slope $-c/{6r}$ in the FTS regime, in which $RD^{r\kappa}$ is large and hence $XZ^{-1/r\nu}=gR^{-1/r\nu}\ll X\ll1$ and $YZ^{-1/r}=L^{-1}R^{-1/r}\ll Y\ll1$ as well as $Z^{-1/r\kappa}=D^{-1}R^{-1/r\kappa}\ll1$ consistently. On the other hand, $f_{SD}$ versus $\ln(L^{-1}D^{\kappa})$ is a horizontal line in the FES regime but changes to an inclined line of a slope $-c/6$ in the FSS regime in which $XY^{1/\nu}=gL^{1/\nu}\ll1$ and $Y^{-1/\kappa}=RL^r\ll1$ as well as $ZY^{-r}=D^{-1}L^{1/\kappa}\ll1$ from Eq.~(\ref{sdtol}).
For the FTS, $f_{SR}$ versus $\ln(D^{-1}R^{-1/r\kappa})$ is a horizontal line in the FTS regime but changes to an inclined line of a slope $-c\kappa/6$ in the FES regime from Eq.~(\ref{srtod}). Similarly, $f_{SL}$ versus $\ln(D^{-1}L^{1/\kappa})$ is a horizontal line in the FSS regime but changes to an inclined line of a slope $-c\kappa/6$ in the FES regime from Eq.~(\ref{sltod}). All these leading slopes are summarized in Table \ref{charac}.
\begin{table}
\caption{\label{charac} Leading slopes in the FTS, FES, and FSS regimes.}
\begin{ruledtabular}
\begin{tabular}{cllll}
ordinate&abscissa                   &FTS            &FES                 &FSS\\
\hline
$f_{SD}$&$\ln(RD^{r\kappa})$        &$-\frac{c}{6r}$&\quad$0$            &\quad$/$\\
$f_{SD}$&$\ln(L^{-1}D^{\kappa})$    &\quad $/$      & \quad $0$          &$-\frac{c}{6}$\\
$f_{SR}$&$\ln(D^{-1}R^{-1/r\kappa})$&\quad$0$       &$-\frac{c\kappa}{6}$& \quad $/$\\
$f_{SR}$&$\ln(L^{-1}R^{-1/r})$      &\quad $0$      & \quad$/$           &$-\frac{c}{6}$\\
$f_{SL}$&$\ln(RL^{r})$              &$-\frac{c}{6r}$&\quad$/$            &\quad$0$\\
$f_{SL}$&$\ln(D^{-1}L^{1/\kappa}) $ & \quad$/$      &$-\frac{c\kappa}{6}$&\quad$0$\\
\end{tabular}
\end{ruledtabular}
\end{table}

The above discussion of the various regimes and their crossovers resembles that of just two scales in Refs.~\cite{Liu,Huang,Hu}. We have just considered pairs of the scales. The other two scales were assumed to be long enough to serve only as perturbations. These perturbations can, however, affect the accuracy of a numerical study.

To be accurate, we thus choose $g=0$, the critical point, fix one scaled parameter to a constant, and vary the remaining one. Interesting behavior can emerge even in this restricted case. For example, upon fixing $Y=L^{-1}D^\kappa=Y_0$, a constant, Eq.~(\ref{sdtor}) is reduced to
\begin{equation}
f_{SD}(0,Y_0,Z)=\frac{c}{6}\ln Z^{-\frac{1}{r}}+ f_{SR}(0,Z^{-\frac{1}{r\kappa}},Y_0Z^{-\frac{1}{r}}).\label{sdtorr}
\end{equation}
In Eq.~(\ref{sdtorr}), $Z=RD^{r\kappa}$ on the two sides is reciprocal and corresponds to two regimes. If $Y_0\ll1$, on the one hand, from the above discussion, it is no double that when $Z$ is vanishingly small, $f_{SD}$ versus $\ln Z=\ln(RD^{r\kappa})$ is a horizontal line in the FES regime and changes to an inclined line of a slope $-c/{6r}$ as $Z$ becomes sufficiently large in the FTS regime. If $Y_0$ is large, on the other hand, we have $L<D^{\kappa}\sim\xi_D$. As a result, for $Z\ll1$, the shortest length scale is $L$ and the regime is in fact the FSS regime. However, as $Y$ is fixed, the regime still exhibits a horizontal line, which is in fact the projection of the cut at the $Y=Y_0$ plane onto the $Y=0$ plane. It can therefore be expected that the value of $Y_0$ affects the crossover between the regime and the FTS regime. As we have scaled $f_{SD}$ by $D$, we refer the regime as the (apparent) FES regime throughout, though $\xi_D$ may not be the shortest length scale. In order to distinguish the FES and the FSS regimes, we can fix $Z=RD^{r\kappa}=Z_0$ and vary $Y=L^{-1}D^\kappa$. Then $f_{SD}$ versus $\ln Y=\ln(L^{-1}D^\kappa)$ is a horizontal line in the FES regime and a line with a slope of $-c/6$ in the FSS regime for large $Y$. Note, however, that the FES regime can in fact be the FTS regime for a large $Z_0$ too.

\subsection{\label{specialloci}Special loci in the scaling of entanglement entropy}
From the general theory in \ref{entropy}, we can also derive some special loci that follow simple laws.

From Eq.~(\ref{stod}), if one lets $RD^{r\kappa}=Z_0$ be a constant and $L$ infinite long, the scaling function varies with $gD^{\kappa/\nu}$ solely.
Accordingly, at $g=0$, Eq.~(\ref{stod}) becomes
\begin{equation}
S(0,D^{-1},0,R)=\frac{c}{6}\ln D^{\kappa}+f_{SD}(0,0,Z_0).\label{stod00}
\end{equation}
This means that, for a given $Z_0$ and a series of $D$, we can choose a corresponding series of $R$ that satisfies $RD^{r\kappa}=Z_0$ and then vary $g$ via $g=Rt$. At $g=0$, $S$ must obey Eq.~(\ref{etod}) because $f_{SD}$ is identical for all the chosen $D$ and $R$. Note that in this case $g$ is changing while Eq.~(\ref{etod}) was originally proposed in the statics. Therefore, the slope of $S$ versus $\ln D$ gives $c\kappa/6$. Note, however, that this cannot be regarded as a measurement of the exponent because we have let $RD^{r\kappa}=Z_0$. One can of course choose rates fulfilling $RD^{r\kappa}\ll1$ so that this term can be ignored. Yet, this introduces small errors. In addition, because $RD^{r\kappa}$ has been fixed, the fixed constant $Z_0$ can be either small or large, even though in the latter case, the system lies in the FTS regime in which the leading behavior is determined by $R$, Eq.~(\ref{stor}), and $D$ is only a perturbation.

We will see in Sec.~\ref{numr} below that the entanglement entropy always increases rapidly as the system approaches the critical point and oscillates beyond. This offers another special locus, the peak of the entanglement entropy. At the peaks, the derivative of $f_{SD}$ with respect to $X=gD^{\kappa/\nu}$ is zero. Accordingly~\cite{Zhong1,Zhong2,Yin},
\begin{equation}
g_p=c_1D^{-\kappa/\nu},
\label{exp1}
\end{equation}
where the subscript $p$ denotes $g$ at the peak and $c_1$ is a constant satisfying $\partial f_{SD}(X,0,Z_0)/\partial X|_{X=c_1}=0$. 

Similarly, according to Eq.~(\ref{stor}), for $D^{-1}R^{-1/r\kappa}=Y_0$, a constant, and $L\rightarrow \infty$, the entanglement entropy obeys Eq.~(\ref{etor}) at $g=0$ because Eq.~(\ref{stor}) is now
\begin{equation}
S(0,D^{-1},0,R)=\frac{c}{6}\ln R^{-\frac{1}{r}}+f_{SR}(0,Y_0,0),\label{stor00}
\end{equation}
whereas, at the peaks
\begin{equation}
g_p=c_2R^{1/\nu r},
\label{exp2}
\end{equation}
with $c_2$ satisfying $\partial f_{SR}(X,Y_0,0)/\partial X|_{X=c_2}=0$. As pointed out above, $Y_0$ can be either small or large because $D^{-1}R^{-1/r\kappa}$ is fixed.  


\section{\label{model}Model and algorithm}
To confirm the scaling theory, we study the one-dimensional transverse-field Ising model, whose Hamiltonian is~\cite{Sachdev}
\begin{equation}
\mathcal{H}_{\rm I}=-\sum_{i=1}^{L-1} \sigma_{i}^z\sigma_{i+1}^z-h_{x}\sum_{i=1}^L\sigma_i^x,
\label{modelI}
\end{equation}
where $\sigma_{i}^x$ and $\sigma_{i}^z$ are the Pauli matrices at site $i$ in the $x$ and $z$ directions, respectively, and $h_{x}$ is the transverse field in the $x$ direction. We have chosen the coupling constant as the energy unit and set it to unity. The critical point of model (\ref{modelI}) lies at $h_{x}=h_{xc}=1$ and its critical exponents are exactly $\beta=1/8$, $\nu=1$, and $z=1$. Besides, the central charge is $c=1/2$. As a result, $\kappa\approx 2.03425$ from Eq.~(\ref{kappa}). Also,
\begin{equation}
g=h_x-h_{xc}.\label{g}
\end{equation}
All the exponents, including their pertinent combinations, are listed in Table~\ref{charac2}. An experimental realization of the model is found in CoNb$_2$O$_6$~\cite{Coldea}.
\begin{table}
\caption{\label{charac2} Exact critical exponents and their combinations of the one-dimensional transverse field Ising model.}
\begin{ruledtabular}
\begin{tabular}{ccccc}
$c$      &$\kappa$       &$\nu$ &$z$ &$r$  \\
\hline
$0.500$  &$2.034$   &$1.000$ &$1$ &$2.000$ \\
\hline
$c/6$& $\kappa/\nu$    &$c\kappa/6$       &$1/\nu r$   &$c/6r$\\
\hline
$0.083$&  $2.034$      &$0.170$             &$0.500$             &$0.042$\\
\end{tabular}
\end{ruledtabular}
\end{table}

To study the dynamic scaling of the entanglement for a finite system, we employ the TEBD algorithm \cite{Vidal, Vidalitebd} with an open boundary condition. Its basic idea is to expand the wave function into a matrix product form via the Vidal decomposition~\cite{Hatano}. Accordingly, each site is attached to a matrix, which is then updated upon being acted by a local evolution operator, the Suzuki-Trotter decomposition of $\exp(-i\mathcal{H}t)$. We choose the time interval to be $0.01$ and keep our results to three decimal places. Smaller time intervals were found to yield identical results within our precisions.

\begin{figure}[b]
	\centerline{\epsfig{file=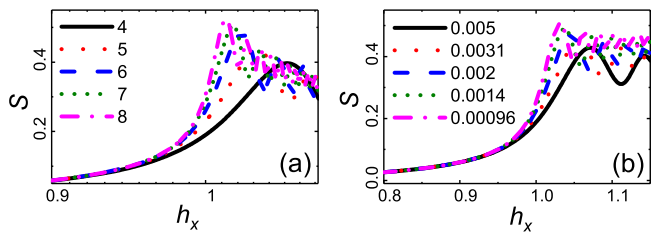,width=1.0\columnwidth}}
  \caption{\label{dynamic} (Color online) $S$ versus $h_x$ with different $D$ and $R$ indicated but (a) fixed $RD^{r\kappa}=0.985$ and (b) fixed $D^{-1}R^{-1/r\kappa}=0.092$, respectively, for an infinitely long Ising chain using its exact critical point and critical exponents.}
  \end{figure}
To calculate the evolution of the entanglement entropy under an external driving for a finite one-dimension system with a finite truncation $D$, we start with the ground state of the Hamiltonian represented in the given $D$ for a sufficiently large $g_i$, which we take as $g_i=-0.7$, corresponding to the initial value of the transverse field $h_{x0}=0.3$. Smaller $g_i$ has been checked to produce negligible differences. Then we increase $g$ linearly with time with a given $R$ until the transition completes. This is just a similar process of a parallel study of the order parameter, in which $g$ is also increased linearly but with a fixed considerably larger $D$~\cite{Yin}.

\section{\label{numr}Numerical results}
In this section, we first provide numerical evidences for the special loci and the scaling forms of the entanglement entropy using iTEBD algorithm for infinite chains~\cite{Vidalitebd}. Then we show the numerical results for the crossovers of the entanglement entropy at the critical point using TEBD algorithm for finite systems with open boundary conditions. In both cases, one scaled variable is fixed to a constant and one is set zero.

\begin{figure}
 \centerline{\epsfig{file=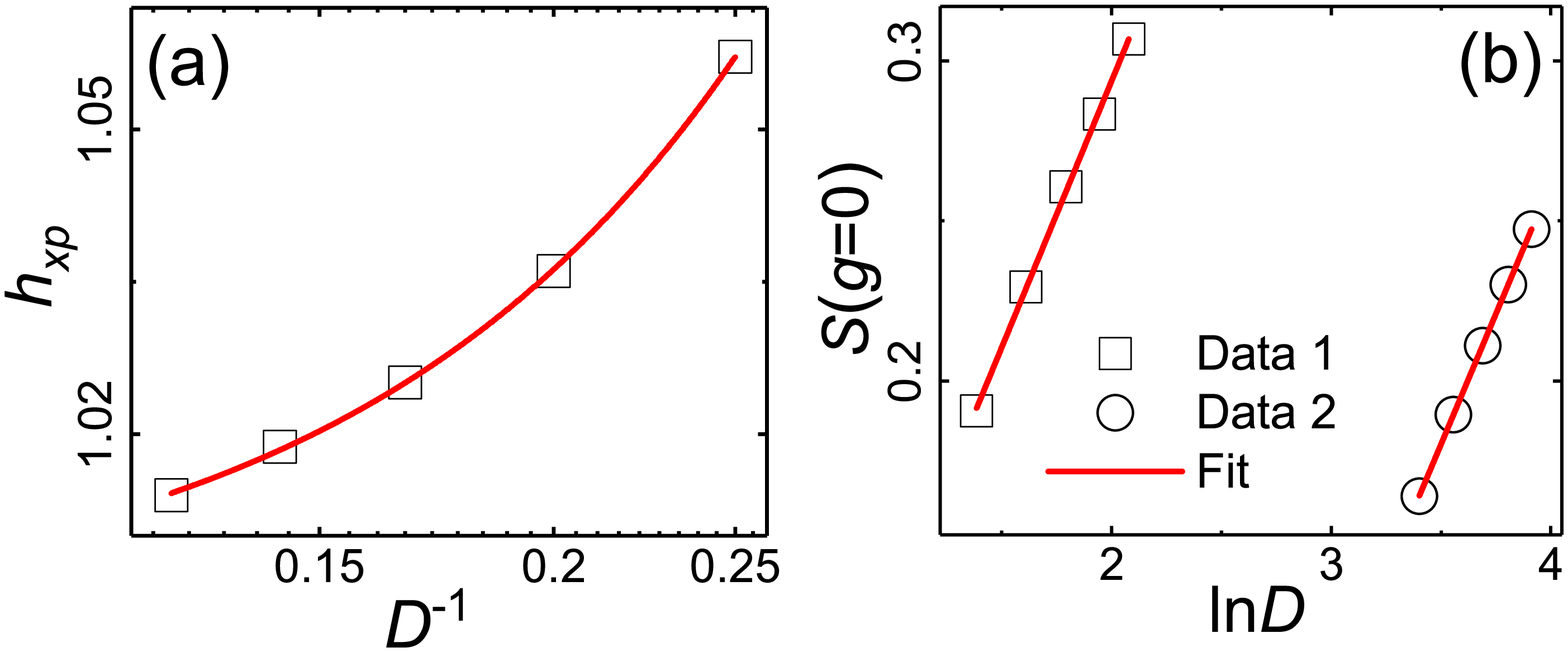,width=1.0\columnwidth}}
  \caption{\label{estimate} (Color online) (a) Fitting of $h_{xp}$ at the peaks of $S$ in Fig.~\ref{dynamic}(a) for the critical point $h_{xc}$ and the critical exponent $\kappa/\nu$ according to Eq.~(\ref{exp1}). (b) Fitting of $S(g=0)$  in Fig.~\ref{dynamic}(a) for the critical exponent $c\kappa/6$ according to Eq.~(\ref{stod00}). Circles represent data for $RD^{r\kappa}=9202.6$.}
\end{figure}
\begin{figure}[b]
 \centerline{\epsfig{file=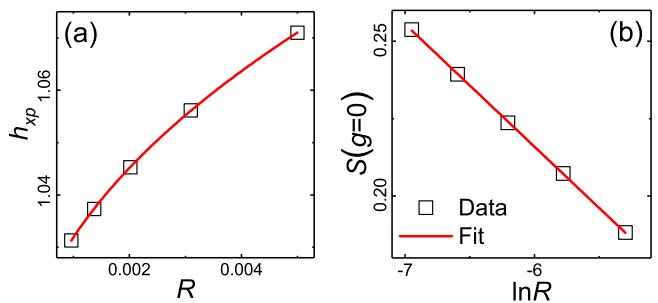,width=1.0\columnwidth}}
  \caption{\label{estimate2} (Color online) (a) Fitting of $h_{xp}$ at the peaks of $S$ in Fig.~\ref{dynamic} (b) for the critical point $h_{xc}$ and the critical exponent $1/\nu r$ according to Eq.(\ref{exp2}). (b) Fitting of $S(g=0)$ in Fig.~\ref{dynamic}(b) for the critical exponent $-c/6r$ according to Eq.~(\ref{stor00}).}
\end{figure}
The entanglement entropy is shown in Fig.~\ref{dynamic}(a) and (b) for various $D$ and $R$, respectively, but fixed $RD^{r\kappa}$. It rises as the transverse field approaches its critical value to a peak beyond $h_{xc}=1$ and then oscillates. It is seen that as $D$ increases and $R$ decreases and thus their corresponding time scales increase, the peaks move closer to the critical point and become higher. Fitting them to Eqs.~(\ref{exp1}) and (\ref{exp2}) with (\ref{g}) as shown in Figs.~\ref{estimate}(a) and \ref{estimate2}(a) yield the critical point at $h_{xc}=1.001$ and $h_{xc}=0.998$ and the critical exponent $\kappa/\nu =2.089$ and $1/\nu r=0.481$, respectively, all in fairly good agreement with the theoretical results in Table~\ref{charac2}. Also, fitting $S$ at $g=0$ to Eqs.~(\ref{stod00}) and (\ref{stor00}) as demonstrated in Figs.~\ref{estimate}(b) and \ref{estimate2}(b) leads to the critical exponents $c\kappa/6=0.163$ and $c/6r=0.040$, respectively. The relatively large error of less than five percents of the latter may be attributed to the rapid variation of the entanglement entropy at the critical point, as can be seen in Fig.~\ref{dynamic}. In Fig.~\ref{estimate}(b), we have also shown another line for a large $RD^{r\kappa}$. The two lines are parallel and thus confirming the validity of Eq.~(\ref{stod00}) for both large and small fixed $RD^{r\kappa}$ values. These show the consistency of the theory.

\begin{figure}[t]
\centerline{\epsfig{file=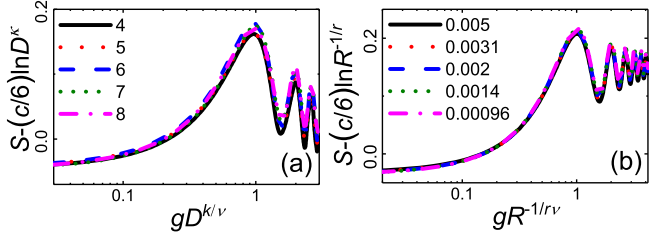,width=1.0\columnwidth}}
  \caption{\label{rescaleddynamic} (Color online) Scaling of the entanglement entropy by (a) $D$ and (b) $R$ for the corresponding curves in Fig.~\ref{dynamic}(a) and (b), respectively.}
  \end{figure}

Moreover, because one scaled variable is fixed in an infinite long system, only one variable is left. We can then scaled the whole entanglement entropy according to Eq.~(\ref{sscaling}). Figure~\ref{rescaleddynamic} shows clearly that the curves in Figs.~\ref{dynamic}(a) and (b) do collapse onto each other well, confirming the FES and the FTS forms, respectively, of the entanglement entropy.

Now we focus on the interesting competition among the time scales, $\zeta_D$, $\zeta_R$, and $\zeta_L$, in a finite system. As pointed out in Sec.~\ref{entropy}, we only study the behavior of the entanglement entropy at the critical point $g=0$ and fix one scaled variable. We demonstrate first the simple crossover between two regimes, each of which is controlled by a single dominant scale and then more complex crossovers involving more scales.

\begin{figure}[b]
  \centerline{\epsfig{file=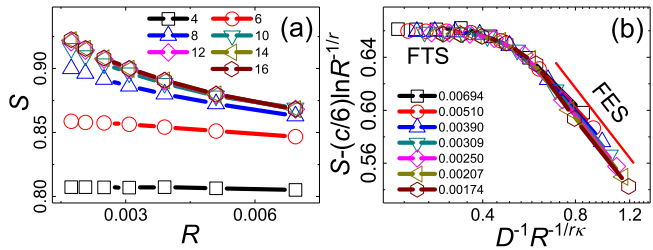,width=1.0\columnwidth}}
  \caption{\label{logR} (Color online) (a) $S$ versus $R^{-1}$ at the critical point $g=0$ for fixed $L^{-1}R^{-1/r}=0.1$. (b) $S-(c/6)\ln R^{-1/r}$ versus $D^{-1}R^{-1/\kappa r}$ for the data in (a) according to Eq.~(\ref{stor}). The two different regions of FTS and FES are marked. The (red) line segment near FES depicts the slope of the leading FES regime according to the exact values listed in Table~\ref{charac2}. The legends in (a) and (b) give the used $D$ and $R$, respectively. Lines connecting symbols are only a guide to the eye.}
\end{figure}
\begin{figure}[b]
  \centerline{\epsfig{file=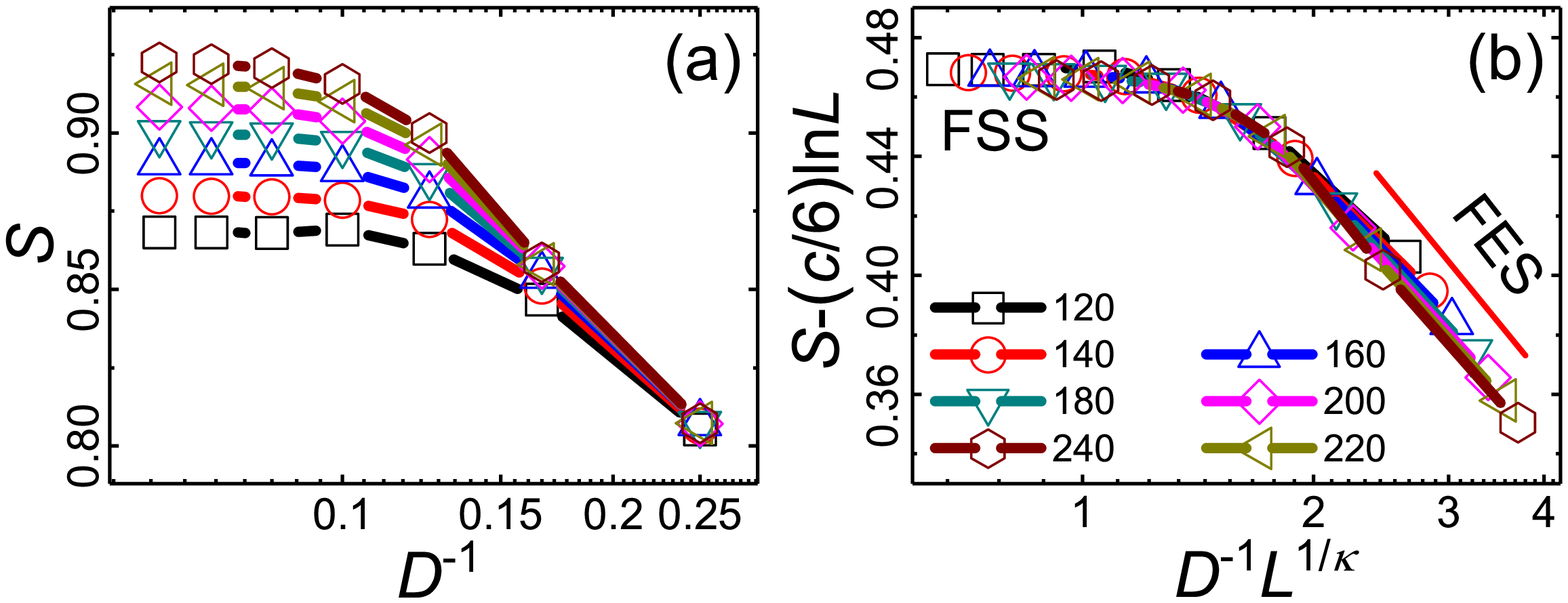,width=1.0\columnwidth}}
  \caption{\label{logL} (Color online) (a) $S$ versus $D^{-1}$ for various $L$ given in (b) at the critical point $g=0$ for fixed $RL^r=100$. (b) $S-(c/6)\ln L$ versus $D^{-1}L^{1/\kappa}$ for the data in (a) according to Eq.~(\ref{stol}). The two different regions of FSS and FES are marked. The (red) line segment near FES depicts the slope of the leading FES regime according to the exact values listed in Table~\ref{charac2}. Lines connecting symbols are only a guide to the eye.}
\end{figure}
We start with the simple crossover between two dominating regimes of FTS and FES by scaling the data with the rate $R$, or in the FTS form~\cite{Huang}, using Eq.~(\ref{stor}). We set $L^{-1}R^{-1/r}=0.1$ so that $L\gg\xi_R$ or $\zeta_L\gg\zeta_R$ and the lattice size can be ignored. From Fig.~\ref{logR}(a), one sees that for large bond dimension $D$, the system falls in the FTS regime where the entanglement entropy $S$ is nearly independent on $D$ but changes with $R$; whereas, for the small $D$, the system enters the FES regime and the dependences reverse: $S$ hardly depends on $R$ but changes with $D$. These are confirmed in Fig.~\ref{logR}(b) upon scaling the data by $R$. On the one hand, for $D^{-1}R^{-1/r\kappa}\ll1$, namely $\zeta_D\gg\zeta_R$, the system is in the FTS regime and the scaled value is almost a constant, which conforms with the characterization of Eq.~(\ref{stor}). On the other hand, for $D^{-1}R^{-1/r\kappa}\gg1$, the system evolves to the FES regime with a slope of $0.164$ close to the exact value $c\kappa/6$ in Table~\ref{charac2} in agreement with Eq.~(\ref{srtod}).

Next, we show the simple crossover between two regimes of FSS and FES. What is distinctive from the above case is that the FSS regime is now controlled by a second shortest rather than by the usual shortest scale. To this end, it is instructive to present the above results in the FSS form to study the competition between $\zeta_L$ and $\zeta_D$ on the basis of Eq.~(\ref{stol}). This is possible because fixing $L^{-1}R^{-1/r}=0.1$ in the FTS form is equivalent to fixing $RL^r=100$ for $r=2$. Although this fixed number is large rather than small as is implied in Eq.~(\ref{stol}), we will see shortly that it is still valid, which validates the logic behind Eqs.~(\ref{xicross}), (\ref{xicrossi}) and (\ref{scross}). Figure~\ref{logL} is just Fig.~\ref{logR} but in the FSS form of representation. One sees from Fig.~\ref{logL}(a) that for large $D$, $S$ depends on $L$ but hardly on $D$, a saturated phenomenon for the dynamic case~\cite{Tagliacozzo}. This is the behavior of FSS, though, it is in fact the FTS regime since the chosen number implies $\zeta_L\gg\zeta_R$. On the other hand, for small $D$, the system enters the FES regime and the dependences on $L$ and $D$ reverse. All data collapse onto a single curve correctly after being scaled in Fig.~\ref{logL}(b). The scaled curve exhibits the two regimes for $D^{-1}L^{1/\kappa}$ small and large, respectively, with the slope of the FES regime equal to $-0.167$ in agreement with $c\kappa/6$ from Tables~\ref{charac} and \ref{charac2}. Therefore, although $L$ is longer than $\xi_R$, the system still shows an apparent FSS regime, which can then cross over to the FES regime as if it were a real FSS regime.

\begin{figure}
  	\centerline{\epsfig{file=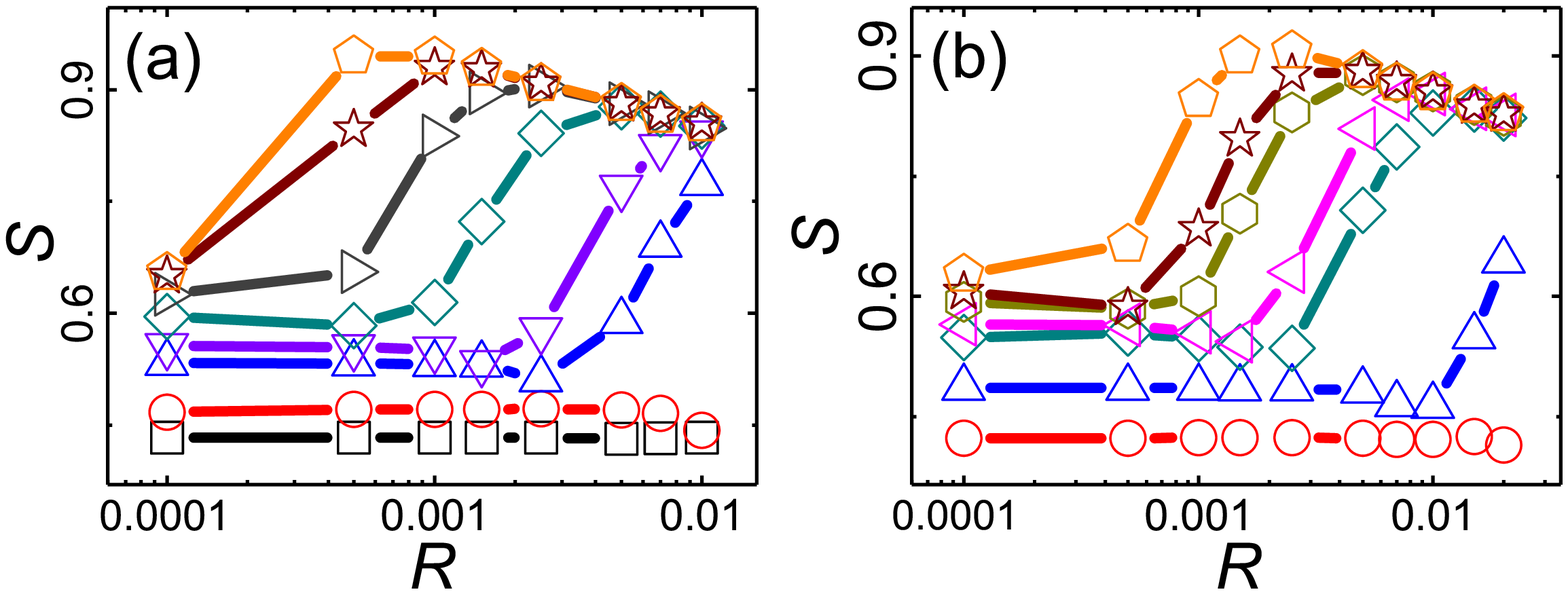,width=1.0\columnwidth}}
\smallskip
\centerline{\epsfig{file=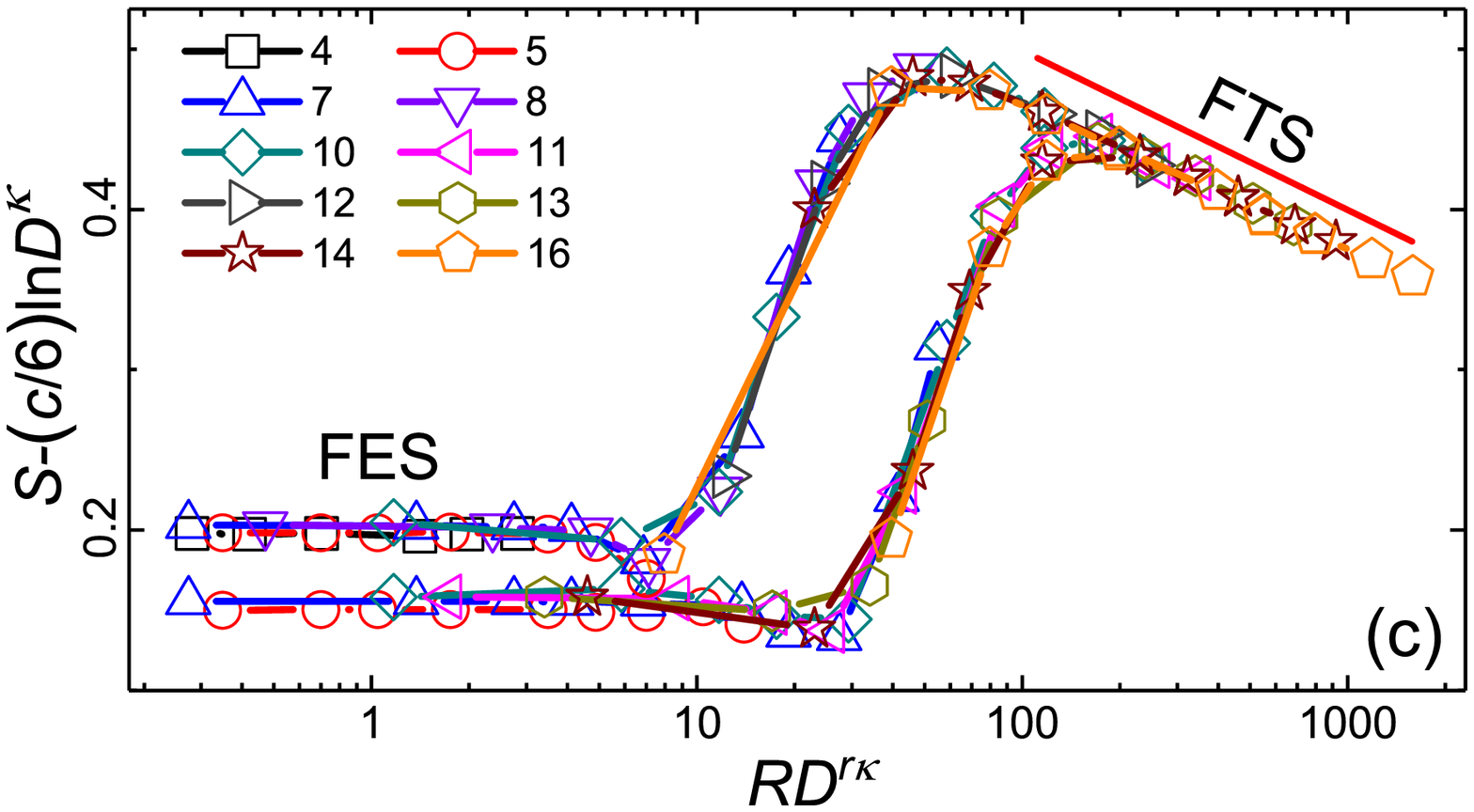,width=1.0\columnwidth}}
  \caption{\label{logDR} (Color online) $S$ versus $R$ for (a) $L^{-1}D^{\kappa}= 0.93$ and (b) $L^{-1}D^{\kappa}= 1.65$, respectively, and various $D$ given in (c) at the critical point $g=0$. (c) $S-(c/6)\ln D^{\kappa}$ versus $RD^{r\kappa}$ according to Eq.~(\ref{stod}) for all data in (a) and (b), corresponding to the left and the right curves, respectively.  The two different regimes of the FES and FTS are marked. The (red) line segment near FTS shows the slope of the leading FTS regime according to the exact values listed in Tables~\ref{charac} and \ref{charac2}. Lines connecting symbols are only a guide to the eye. Semilogarithmic scales are used.}
\end{figure}
\begin{figure}
\centerline{\epsfig{file=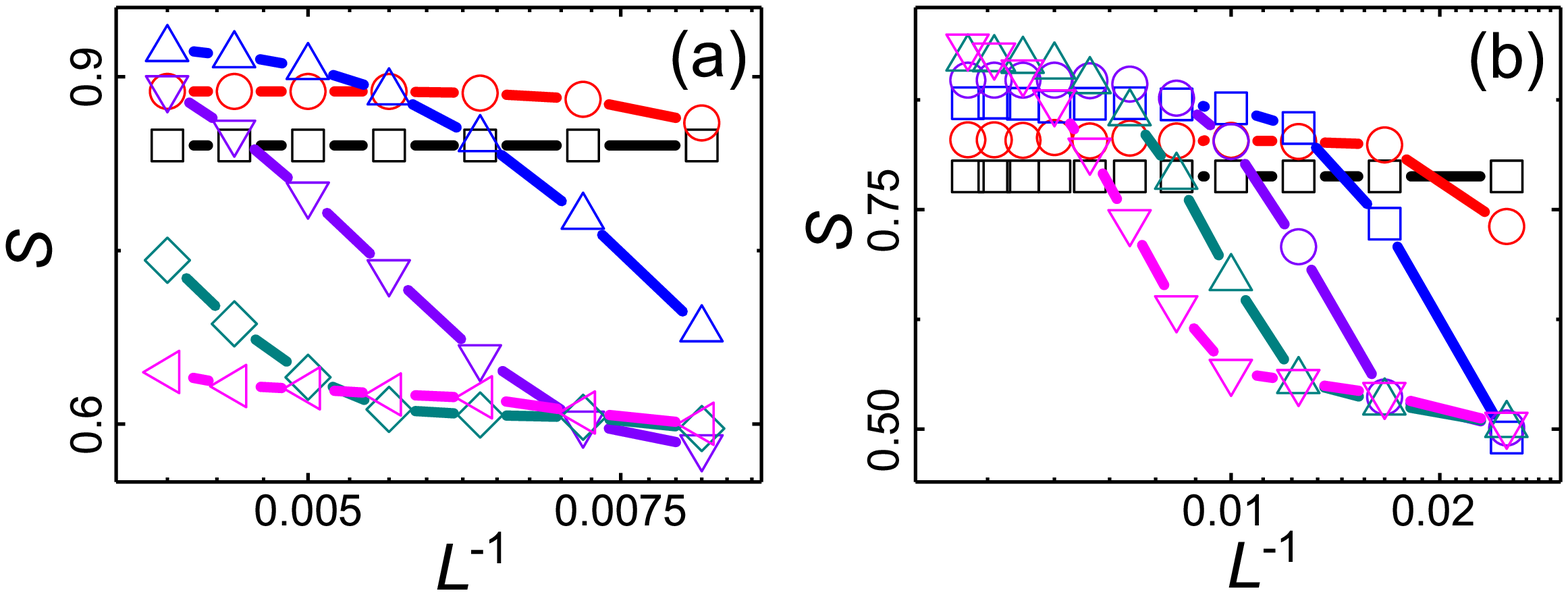,width=1.0\columnwidth}}
\smallskip
\centerline{\epsfig{file=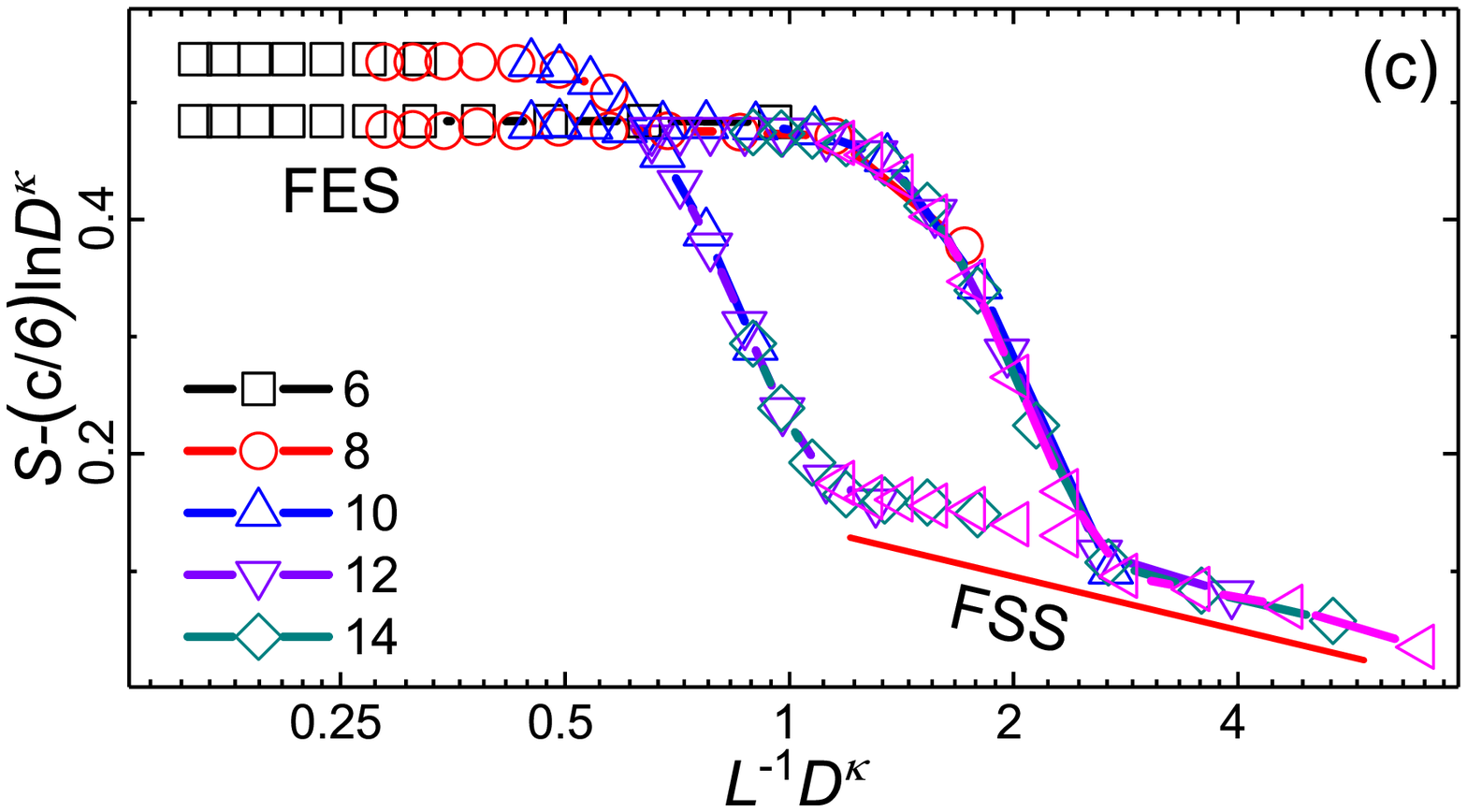,width=1.0\columnwidth}}
  \caption{\label{logD} (Color online) $S$ versus $L^{-1}$ for (a) $RD^{r\kappa}=14.075$ and (b) $RD^{r\kappa}=80$, respectively, and various $D$ given in (c) at the critical point $g=0$. (c) $S-(c/6)\ln D^{\kappa}$ versus $L^{-1}D^{\kappa}$ according to Eq.~(\ref{stod}) for all data in (a) and (b), corresponding to the left and the right curves, respectively. The two different regimes of the FES and FSS are marked. The (red) line segment near FSS shows the slope of the leading FSS regime according to the exact values listed in Tables~\ref{charac} and \ref{charac2}. Lines connecting symbols are only a guide to the eye. Semilogarithmic scales are used.}
\end{figure}
Finally, we show how the fixed term can also lead to complicated behavior. Firstly, we fix the term $L^{-1}D^\kappa$ in $f_{SD}$ to a constant to study the crossover between FES and FTS by scaling the data with $D$ in contrast with Fig.~\ref{logR}(b), or in the FES form, using Eq.~(\ref{stod}). In Figs.~\ref{logDR}(a) and (b), we show the dependence of the entanglement entropy $S$ on $R$ for various $D$ for two different fixed $L^{-1}D^{\kappa}$ values. Comparing with the simple power-law relation in Fig.~\ref{estimate2}(b) or even the competition of two scales in Fig.~\ref{logR}(a), one sees that it appears complicated without exhibiting any simple relation. This is because we have deliberately chosen proper ranges of $R$, $D$, and $L$ so that the three scales $\zeta_R$, $\zeta_D$, and $\zeta_L$ are comparable. Nevertheless, after being scaled, the data again collapse well onto two single curves as demonstrated in Fig.~\ref{logDR}(c), confirming Eq.~(\ref{stod}). From Fig.~\ref{logDR}(c), it can be seen that, for large $RD^{r\kappa}$, $\zeta_R\ll\zeta_D$ and the system lies in the FTS regime. Both the left and the right curves fit on a single curve described by the same function $f_{SD}$ or $f_{SR}$ at the same limit. Indeed, their slopes are $-0.043$ and $-0.040$, respectively, which agree well with the theoretical slope of $-c/6r$, illustrated by the red line, as given in Tables~\ref{charac} and \ref{charac2}. The small difference arises from the ranges used in the fits. However, for small $RD^{r\kappa}$, the regime marked FES exhibits two different horizontal lines, with the smaller the $L^{-1}D^{\kappa}$ value, the further the FTS regime and the higher the value of the horizontal line. This is because only for small $L^{-1}D^{\kappa}$ does the system fall in the true FES regime. For large $L^{-1}D^{\kappa}$, $L\ll\zeta_D$ and the system settles in fact in the FSS instead of the FES regime, similar to what Fig.~\ref{logL}(b) has shown. In between these two FES and FSS regimes, there exists a crossover between them with a slope $-c/6$ from Table~\ref{charac}. Accordingly, within this crossover regime, as $L^{-1}D^{\kappa}$ gets larger, the plane at this value cuts the surface of $f_{SD}(0, L^{-1}D^{\kappa}, RD^{r\kappa})$ at a lower value. In addition, for the same $D$, a larger $L^{-1}D^{\kappa}$ means a shorter $L$ and thus one must use a bigger $R$ in order to be in the FTS regime. This moves the crossover between the apparent FES regime and the FTS regime to a larger $RD^{r\kappa}$ value. This in turn reflects the boundary of $L^{-1}D^{\kappa}\sim (RD^{r\kappa})^{1/r}$ between the FSS and FTS regimes. Note the sharp difference of the crossover from those in Figs.~\ref{logR}(b) and \ref{logL}(b). From the trend of the two displayed curves in Fig.~\ref{logDR}(c), it can be expected that the crossover may be similar to those in Figs.~\ref{logR}(b) and \ref{logL}(b)  when $L^{-1}D^{\kappa}$ are very small or very large so that either $L$ or $D$, respectively, can be neglected. In between, the three scales conspire to produce the feature. Therefore, complicated regimes and crossovers appear when the three scales are involved. Nevertheless, they can be understandable and describable by the scaling theory.

To further corroborate the above picture, in Fig.~\ref{logD}(c), we demonstrate the projection onto the other $RD^{r\kappa}=0$ plane the collapses of $S$ using Eq.~(\ref{stol}) for its dependence on $L^{-1}$ for various $D$ at two fixed $RD^{r\kappa}$ equal to $14.075$ and $80$ shown in Fig.~\ref{logD}(a) and (b), respectively. From Fig.~\ref{logD}(c), for large $L^{-1}D^{\kappa}$ we see now that both the left and the right curves coincide in this FSS dominated regime. Their slopes are $-0.074$ and $-0.078$, respectively, in agreement with $-c/6$ from Tables~\ref{charac} and \ref{charac2}. However, for small $L^{-1}D^{\kappa}$, the projections of the two cuts on the two $RD^{r\kappa}$ planes onto the $RD^{r\kappa}=0$ plane again lead to two horizontal lines marked as the FES regime too, though in both cases of large $RD^{r\kappa}$ values, the apparent FES regime is in fact the FTS regime. In consistence with Fig.~\ref{logDR}(c), the smaller the $RD^{r\kappa}$ value, the further the FSS regime, and the higher the value of the horizontal line in the FES regime. Also the special crossovers reflect again the crossover between the FSS and the FTS regimes.

\section{\label{sum}Summary}
We have studied the scaling behavior of the entanglement entropy under several confinements near the critical point. These confinements bring to the system several corresponding time or length scales besides its intrinsic correlation time or length. A unified scaling theory of the entanglement entropy for the competition of these scales has been set up and verified via the one-dimensional transverse field Ising model with varied bond dimensions and subjected to a time-dependent driving in a finite size system. According to the theory, the system is dominated by the shortest among the scales when the other scales are sufficiently longer than that scale. As a consequence, the system can exhibit FTS, FES, FSS, and (quasi-)equilibrium scaling once the driving length scale, the bond-dimension length scale, the system size, and the equilibrium correlation length are the respective governing scale. Crossover takes place when the leading scale changes from one scale to another one. When one scaled variable is fixed so that the relative lengths of the two scales comprising the scaled variable are fixed, interesting behavior appears. In this case, a second shortest scale can possess its own regime since the shortest scale is bound up with it. Nevertheless, the shortest scale can contribute subtly to the crossover behavior between the second shortest and yet another scale, especially when the three involved scales are of comparable sizes. We believe that the theory and its results are instructive to experiments when more than two scales have to be considered.

\section*{Acknowledgement}
This work was supported by the National Natural Science Foundation of China (Grant No. 11575297).

\end{document}